# Linear-Model-inspired Neural Network for Electromagnetic Inverse Scattering

Huilin Zhou, Tao Ouyang, Yadan Li, Jian Liu, Qiegen Liu, *Senior Member, IEEE*

*Abstract*—**Electromagnetic inverse scattering problems (ISPs) aim to retrieve permittivities of dielectric scatterers from the scattering measurement. It is often highly nonlinear, causing the problem to be very difficult to solve. To alleviate the issue, this letter exploits a linear model-based network (LMN) learning strategy, which benefits from both model complexity and data learning. By introducing a linear model for ISPs, a new model with network-driven regularizer is proposed. For attaining efficient end-to-end learning, the network architecture and hyper-parameter estimation are presented. Experimental results validate its superiority to some state-of-the-arts.**

## I. INTRODUCTION

ELECTROMAGNETIC inverse scattering problems (ISPs) are devoted to reconstructing the location, shape, and some electrical properties of unknown objects from the measured scattered fields. ISPs have been attracting attentions for many years and have various applications [1-6], such as nondestructive testing, through-wall imaging, geophysics, and remote sensing. They can be roughly described by Lippmann–Schwinger equation for the field inside and outside the scattering object. It is challenging to address due to the intrinsically ill-posedness and non-linearity property.

Related solutions for ISPs have been vigorously presented, which can be divided into linear and non-linear methods. In the category of linear method, it is assumed that the difference between the incident field and the total field is less than 30%, and thus the first-order Born approximation (BA) is used to describe the total field [7]. In this circumstance, the total field inside the medium with the incident field can be replaced by ignoring the multiple scattering effect between the target and the medium, thereby linearizing the target contrast and receiving that related to the scattering field [8]. Additionally, many regularization methods were proposed to alleviate the ill-posed deficiency, including the truncated singular value decomposition (TSVD) [9] and low rank constraint [10]. In summary, these methods have higher parameter inversion speed and accuracy in the case of the medium target to be a weak scatter, i.e., the difference between the dielectric constant of the target and the background is small. Nonetheless, they are not suitable for strong scatterers. The category of nonlinear method considers the fact that the multiple scattering effects are caused by the interaction of the incident wave with medium target. The total field in the imaging region is the superposition of the incident field and the scattering field. Due to the inherent non-linearity and ill-posedness of ISPs, regularization-induced iterative optimization methods are put forward, like contrast source-type inversion (CSI) [13–14], Born iterative method and its variants [11-12], and sub-space optimization method (SOM) [15–16]. The benefit of iterative strategy is that the formula is rigorous and can re-construct the spatial distribution of strong scatterers. Nevertheless, it is often sensitive to initial values and converges in relatively slow speed. It also needs to manually set the iteration number and regularization parameters. In addition, many methods based on multi-resolution (MR) schemes have been proposed to counteract both non-linearity and ill-posedness of ISPs [17].

In recent years, deep learning has gained promising performance in various engineering applications [21-23]. The artificial neural network has been utilized to resolve the ISPs in [18], [20]. Nevertheless, these methods used a few parameters to represent scatterers, leading to the limitations of this strategy. Later, three inversion schemes based on the U-Net convolutional neural network (CNN) have been proposed in [24]. Meanwhile, the connection between conventional iterative algorithms and DNNs has been profitably investigated in [19], termed DeepNIS. DeepNIS is based on a cascade of three CNN modules, in which the inputs from the backpropagation (BP) method and the outputs are the reconstruction images of the unknown scatterers. Although employing deep learning scheme directly may attain outstanding results, it fails to integrate with the physical knowledge of electromagnetic inverse scattering. Inspired by traditional nonlinear iterative algorithms, several CNN learning strategies that incorporating physical expertise have achieved outstanding inversion accuracy and efficiency [26]. A two-step deep learning approach has been proposed in [25], which can reconstruct high-contrast objects using the cascade of a CNN and another complex-valued deep residual CNN.

The above mentioned deep learning-based inverse scattering research has mainly focused on data-driven and non-linear iterative physical model-driven methods. To the best of our

This work was supported in part by the National Natural Science Foundation of China under 61561034. Author for correspondence: H. Zhou.
H. Zhou, T. Ouyang, Y. Li, J. Liu and Q. Liu are with the Department of Electronic Information Engineering, Nanchang University, 330031, China.

({416114417145, ouyangtao, liujian}@email.ncu.edu.cn, {zhouhuilin, liuqiegen}@ncu.edu.cn).

knowledge, there has no work to tackle the linear problems via deep learning scheme. In this letter, we propose a scheme based on CNN learning to handle the ISPs with linear constraint [27].

## II. LMN: LINEAR MODEL-BASED NETWORK

### A. Linear Model for ISPs

A two-dimensional (2D) transverse magnetic (TM, i.e. Ez polarization) ISPs is shown in Fig. 1. The unknown nonmagnetic scatterers are within a free-space background DOI ($D \subset R^2$). They are illuminated by incoming electromagnetic waves, which are generated by transmitters located at $r_j$, $j = 1, 2, \cdots, N_j$. For each incidence, the scattered field is measured by an array of receivers that located at $r_q$, $q = 1, 2, \cdots, N_s$.

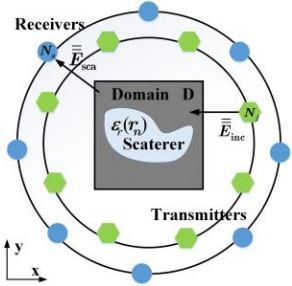

Fig. 1. Schematic diagram of inverse scattering problems.

The forward formulation of ISPs can be described by two equations. Specifically, the first one is the Lippmann–Schwinger equation, or equivalent domain integral equation:

$$\bar{\bar{E}}_{tot} = \bar{\bar{E}}_{inc} + \bar{\bar{G}}_D \bar{\bar{\chi}} \bar{\bar{E}}_{tot} \qquad (1)$$

where $\bar{\bar{E}}_{inc}$ and $\bar{\bar{E}}_{tot}$ denote the incident and total electric fields, respectively. The diagonal matrix $\bar{\bar{\chi}}$ denotes the contrast of reconstructed object scatterer whose diagonal element is $\bar{\bar{\chi}}(n,n) = \varepsilon_r(r_n) - 1$, $\varepsilon_r(r_n)$ is the relative permittivity at $r_n$. $\bar{\bar{G}}_D$ is a 2D free space Green's function in domain $D$.

The discretized formulation of the second equation in ISPs is

$$\bar{\bar{E}}_{sca} = \bar{\bar{G}}_S \bar{\bar{\chi}} \bar{\bar{E}}_{tot} \qquad (2)$$

where $\bar{\bar{E}}_{sca}$ is the scattered field. $\bar{\bar{G}}_S$ is a 2D free space Green's function in domain $S$.

The pursuit of inverse scattering problem is to determine the relative permittivities of the scatterer from the observed scattered field $\bar{\bar{E}}_{sca}$ via Eqs. (1)(2). At a first glance, it is a nonlinear equation. However, as discussed in Introduction section, in some special case, ISPs can be solved by BA [7]. i.e.,

$$\bar{\bar{E}}_{sca} = \bar{\bar{G}}_S diag(\bar{\bar{E}}_{inc}) \bar{\bar{\chi}} \qquad (3)$$

By introducing regularization term into the linear model, there exists

$$\bar{\bar{\chi}} = \arg\min_{\bar{\bar{\chi}}} \left\| \bar{\bar{E}}_{sca} - \bar{\bar{G}}_S diag(\bar{\bar{E}}_{inc}) \bar{\bar{\chi}} \right\|_2^2 + \lambda \| R_w(\bar{\bar{\chi}}) \|^2 \qquad (4)$$

where $R_w(\bar{\bar{\chi}})$ is a learnable regularizer CNN estimator that depends on the network parameters $w$.

### B. Network Architecture of LMN

By setting $R_w(\bar{\bar{\chi}})$ to be network-contained, it yields,

$$R_w(\bar{\bar{\chi}}) = \bar{\bar{\chi}} - V_w(\bar{\bar{\chi}}) \qquad (5)$$

where $V_w(\bar{\bar{\chi}})$ represents the denoising part of $\bar{\bar{\chi}}$, removing the ghost and noise. The CNN-based prior $\| R_w(\bar{\bar{\chi}}) \|^2$ tries to avoid signal corruption as well as guarantee data-consistency. Substituting Eq. (5) into Eq. (4), it attains:

$$\bar{\bar{\chi}}_{rec} = \arg\min_{\bar{\bar{\chi}}} \left\| \bar{\bar{E}}_{sca} - \bar{\bar{G}}_S diag(\bar{\bar{E}}_{tot}) \bar{\bar{\chi}} \right\|_2^2 + \lambda \| \bar{\bar{\chi}} - V_w(\bar{\bar{\chi}}) \|^2 \qquad (6)$$

By introducing auxiliary intermediate variable $Z$, we obtain an alternating iterative formulation that approximates to Eq. (6):

$$\bar{\bar{\chi}}_{n+1} = \arg\min_{\bar{\bar{\chi}}} \left\| \bar{\bar{E}}_{sca} - \bar{\bar{G}}_S diag(\bar{\bar{E}}_{tot}) \bar{\bar{\chi}} \right\|_2^2 + \lambda \| \bar{\bar{\chi}} - Z_n \|^2 \qquad (7a)$$

$$Z_{n+1} = V_w(\bar{\bar{\chi}}_{n+1}) \qquad (7b)$$

By calculating the gradient of sub-problem Eq. (7a) and letting it to be zero, it attains:

$$[(\bar{\bar{G}}_S diag(\bar{\bar{E}}_{tot}))^H (\bar{\bar{G}}_S diag(\bar{\bar{E}}_{tot})) + \lambda I] \bar{\bar{\chi}}_{n+1} = (\bar{\bar{G}}_S diag(\bar{\bar{E}}_{tot}))^H (\bar{\bar{E}}_{sca}) + \lambda Z_n \qquad (8)$$

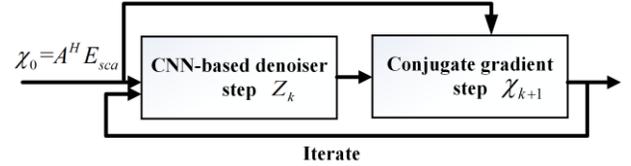

Fig. 2. An illustration of the iteration scheme in Eq. (7b) and Eq. (8).

The schematic diagram of the iterative framework is shown in Fig. 2, where $A = \bar{\bar{G}}_S diag(\bar{\bar{E}}_{tot})$. After initializing with $\bar{\bar{\chi}}_0 = A^H \bar{\bar{E}}_{sca}$, it alternatively updates $Z_n$ and $\bar{\bar{\chi}}_{n+1}$ by CNN-based denoiser step Eq. (7b) and conjugate gradient step Eq. (8).

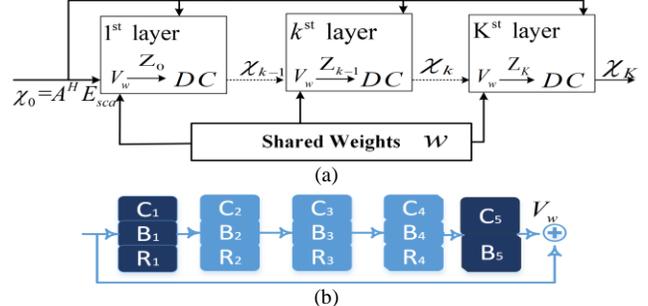

Fig. 3. The flowchart of the LMN network (a) and the CNN denoiser sub-network (b). Here the convolution (Conv), batch normalization (BN) and rectified linear unit (ReLU) layers are denoted as "C", "B" and "R", respectively. The dimension and number of the convolutional filters are $3 \times 3$ and 64, respectively.

By regarding one iteration as one layer, the above update rule can be viewed as an unrolled deep CNN, whose weights at different iterations are shared. An end-to-end training scheme is employed to optimize it. The proposed unrolled architecture uses the same denoising operator $V_w(\bar{\bar{\chi}})$ at each layer, hence significant reduction in model complexity is allowed. Besides, $\lambda$ is a trainable regularization parameter. High $\lambda$-value in the training procedure indicates that the constrained setting can achieve improvement.

In summary, the flowchart of network LMN is depicted in Fig. 3, which consists of a series of learnable CNN denoiser sub-network and data-consistency (DC) sub-network. Specifically, the widely used residual network is adopted in the CNN denoiser sub-network. At the meantime, conjugate gradient is done in the DC sub-network.

### C. Parameter Solver of LMN

As seen in the previous subsection, decoupling the training process from the specifics of the acquisition procedure simplifies the approach. Besides of the network architecture, we leverage the performance of the network via directly minimizing the loss function by means of end-to-end fashion. i.e., assuming the number of layer to be $K$, the loss function between $\bar{\bar{\chi}}_K$ and the desired image $\hat{\bar{\bar{\chi}}}$ is defined as follows:

$$L = \sum_{i=1}^{Nsamples} \left\| \bar{\bar{\chi}}_K(i) - \hat{\bar{\bar{\chi}}}(i) \right\|^2 \quad (9)$$

where $\hat{\bar{\bar{\chi}}}(i)$ is the $i$-th label image. By means of minimizing Eq. (9), the parameters in CNN sub-network and DC sub-network will be updated via training on data pairs.

At first, the gradient of the cost function Eq. (9) with respect to the shared weights $w$ is given by the chain rule

$$(\nabla_w L) = \sum_{k=0}^{K-1} J_w(\mathbf{Z}_k)^T (\nabla_{\mathbf{Z}_k} L) \quad (10)$$

where the element of the Jacobian matrix $J_w(\mathbf{Z})$ is $[J_w(\mathbf{Z})]_{i,j} = \partial z_i / \partial w_j$ and $z_i$ is the output of CNN at $k$-th layer.

In order to apply the backpropagation scheme to Eq. (10), the next step is how to evaluate the terms $\nabla_{\mathbf{Z}_k} L, k = 0, \cdots, K-1$. According to the formulation of Eq. (8), we backpropagate them via designing numerical optimization blocks, i.e., conjugate gradient (CG) blocks [28]. Concretely, according to the chain rule, it has

$$\nabla_{Z_{k-1}} L = J_{Z_{k-1}}(\bar{\bar{\chi}}_k)^T \nabla_{\bar{\bar{\chi}}_k} L \quad (11)$$

where the Jacobian matrix $J_Z(\bar{\bar{\chi}})$ has entries $[J_Z(\bar{\bar{\chi}})]_{i,j} = \partial x_i / \partial z_j$. By calculating the gradient from Eq. (8), the value of Jacobian matrix is given by

$$J_Z(\bar{\bar{\chi}}) = [(\bar{\bar{G}}_s diag(\bar{\bar{E}}_{tot}))^H (\bar{\bar{G}}_s diag(\bar{\bar{E}}_{tot})) + \lambda I]^{-1} \quad (12)$$

Finally, after Eq. (11) is determined, the network parameter of the network $V_w$ and the regularization parameter $\lambda$ can also be updated by Adam optimization scheme [29].

## III. NUMERICAL RESULTS

The performance of proposed network LMN is evaluated in reconstructing relative permittivities from scattered fields. We implemented its architecture in MATLAB on a PC equipped with Inter(R) Core (TM) i7-7800X CPU and GeForce Titan 1080Ti. Results under synthetic data including circular-cylinder and MNIST datasets [30] are presented.

### A. Experiment Configuration

In the experiment, a domain of interest(DOI) with size of 2×2 $m^2$ is divided into 128×128 pixels. In the inversion process, in order to avoid the inverse crime, the DoI is divided into 30×30 and 64×64 pixels, respectively. Among yhem, 16 line sources and 32 line receivers are equally placed on a circle centered at

(0, 0) m and with diameter 12 m and 6 m. The scattered fields are generated numerically using the method of moment (MOM) and recorded into $\bar{\bar{E}}_{sca}$, which is a matrix with $N_r \times N_i$ dimensions. By inserting additive white Gaussian noise $\bar{\bar{n}}$, the measured scattered field $\bar{\bar{E}}_{sca} + \bar{\bar{n}}$ is utilized to reconstruct relative permittivities. The noise level is relatively defined as $\|\bar{\bar{n}}\|_F / \|\bar{\bar{E}}_{sca}\|_F$. The operating frequency is 400 MHz, and a priori information is that the scatterers are lossless as well as fall into the range of nonnegative contrast [13].

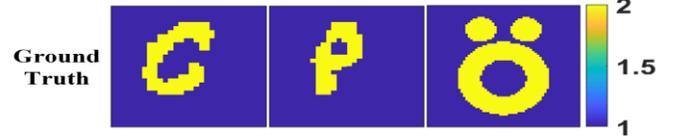

Fig. 4. Ground truth of the test samples

In order to evaluate the reconstruction performance of these algorithms, a relative error of the reconstructed permittivity $R_e$ is defined as

$$R_e = \frac{1}{M_t} \sum_{j=1}^{M_t} \left\| \bar{\bar{\varepsilon}}_r^t - \bar{\bar{\varepsilon}}_r^r \right\|_F / \left\| \bar{\bar{\varepsilon}}_r^t \right\|_F \quad (13)$$

where $\bar{\bar{\varepsilon}}_r^r$ and $\bar{\bar{\varepsilon}}_r^t$ are the reconstructed relative permittivity and ground-truth relative permittivity. $M_t$ is the number of conducted tests.

### B. Algorithm Robustness

In order to investigate the generalization feasibility of the introduced LMN, we thoroughly evaluate the network trained under MNIST dataset [30], which consists of various handwriting Latin letters that used in machine learning community. Rather than recognizing and classifying the Latin letters, we quantitatively reconstruct the profile where the scatterers are represented by the Latin letters. Based on MNIST, the letters-shaped objects are set to be the relative permittivity that belongs to the interval of 1.5-2.4. To ensure the generalization, we randomly choose 200 samples in MNIST dataset for training the network LMN. According to our trail, the performance of the trained models based on the same distribution of randomly chooses samples keeps stable. Another 20 samples are chosen for testing. The profiles in test case are similar from those used in the training phase.

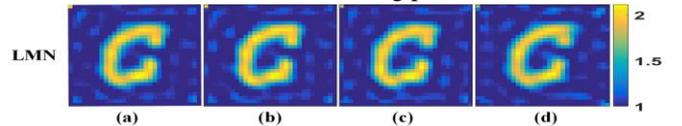

Fig. 5. Relative permittivity distribution reconstructions of LMN from scattered fields under various additive white Gaussian noise. (a)(b)(c)(d) reconstruction results under noise level 0%, 10%, 15%, and 20%.

In the experiment, we use a relative dielectric constant to train the network LMN without noise, and then uses the trained network to predict the test sample at different noise levels, i.e., 0%, 10%, 15%, 20%. Fig. 5 visualizes the reconstructed relative permittivity distributions of some representative examples. It can be observed that LMN is able to reconstruct the profile with promising results. Relative errors of all the reconstructed permittivity and the reconstructed time are listed in Table I. It validates the anti-interference ability of LMN in the environment where the scatterer is in strong noise. In terms of computational efficiency, it is not difficult to find that LMN has realized the effect of real-time reconstruction to a certain extent.

Table I RELATIVE ERRORS and TIME FOR TESTS on "C" FROM MNIST.

|  | 0% | 10% | 15% | 20% | Average Time (s) |
|---|---|---|---|---|---|
| LMN | 0.2837 | 0.2847 | 0.2860 | 0.2848 | 0.6736 |

### C. Comparison under MNIST Database

The effectiveness and superiority of the algorithm is further reflected by comparing the reconstruction results of the two repesentative methods. i.e., the non-iterative algorithm BA and iterative algorithm SOM. LMN is inspired from the first-order Born approximation, and the regularization part is replaced by network learning. In Fig. 6, we reconstruct the scatterers from four different SNR scattering fields. From the pattern reconstructed by the same method BA, as the signal-to-noise ratio decreases, the information loss of the reconstructed object gradually increases. However, the pattern change in the reconstruction of SOM and LMN is obviously small. Additionally, both SOM and LMN reconstruct the scatterers better than BA.

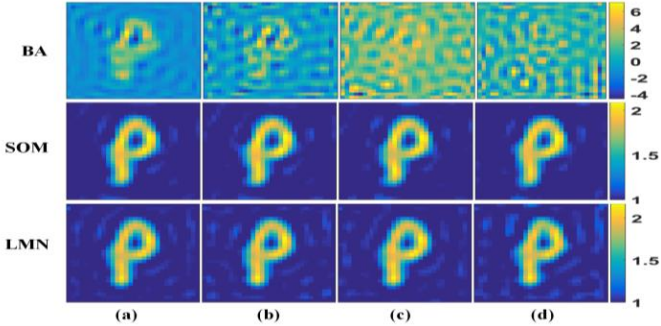

Fig. 6. Reconstructed relative permittivity profiles of BA, SOM, LMN from scattered fields under various noise levels. (a)(b)(c)(d) reconstruction results under noise level 0%, 10%, 15%, and 20%.

Table II RELATIVE ERRORS and TIME FOR TESTS on "p" from MNIST.

|  | 0% | 10% | 15% | 20% | Average Time (s) |
|---|---|---|---|---|---|
| BA | 0.2725 | 0.3161 | 0.3575 | 0.6724 | 445 |
| SOM | 0.2294 | 0.2305 | 0.2338 | 0.2301 | 3.5 |
| LMN | 0.2585 | 0.2592 | 0.2588 | 0.2588 | 0.6921 |

The above analysis is also reflected in Table II. It shows that BA has certain defects. The proposed scheme can minimize the influence of noise on the reconstructed object due to the CNN denoiser, and has good anti-noise performance. Additionally, by observing and analyzing the reconstructed maps of LMN and SOM, it is found that the image quality of SOM is quite near to that of LMN. Although both reconstruction effects are satisfactory, the reconstructed time required in SOM is larger. By comparing LMN with SOM, the superiority of the linear algorithm conbined with data learning over the non-linear algorithm is demonstrated.

### D. Comparison under "Austria profile"

A more classic scatterer is used to test the performance of the proposed solution. The scatterer consists of a ring and two small circles with the same properties. As shown in Fig. 4, the outer circle radius of the ring is 2 m, and the inner circle radius is 1.5 m. The radius of the small circle is 1 m. The relative dielectric constant of the scatterer is 2. The range of relative permittivity is between 1.1 and 2.0 in the noise-free scattered field.

Fig. 7 visualizes the reconstructed relative permittivity profiles for three methods, where 0% ,10% , 20% and 30% Gaussian noises are considered in the scattered fields. Since the additive white Gaussian noise in the test procedure may be much higher than that in the training procedure, the case with 30% noise aomount is investigated in our experiment.

In the case of 30% noise for the first-order Born approximation, due to the large error, the corresponding inverse scattering reconstruction map is not shown here. Specifically, the trained network with only 0% additive white Gaussian noise is ultilized to presented in the training procedure. By observing the reconstruction map, we find that LMN improves the traditional linear method to a certain extent. Although the reconstruction result is only comparable to the non-linear method SOM, LMN uses the same trained network to reconstruct the four different input scattering fields.

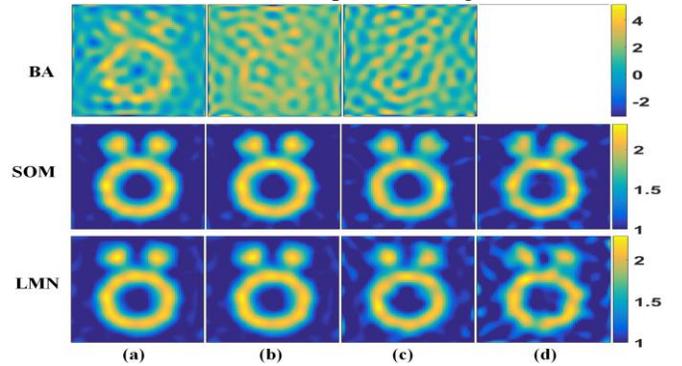

Fig. 7. Reconstructed relative permittivity profiles of BA, SOM, LMN from scattered fields under various noise levels. (a)(b)(c)(d) reconstruction results under noise level 0%, 10%, 20%, and 30%.

Table III RELATIVE ERRORS and TIME FOR THE TESTS on "Austria".

|  | 0% | 10% | 20% | 30% | Average Time (s) |
|---|---|---|---|---|---|
| BA | 0.2077 | 0.4969 | 1.3591 | NA | 875 |
| SOM | 0.1378 | 0.1389 | 0.1421 | 0.1446 | 51 |
| LMN | 0.1432 | 0.1440 | 0.1503 | 0.1602 | 0.3825 |

In Table III, it is found that the relative errors in the reconstruction of the scattered field remain almost unchanged under different noise levels, Therefore, it can be concluded that LMN has a certain anti-interference ability in high noise environment, maintaining good reconstruction effect. In terms of computational times, the time is significantly shorter than the other two algorithms.

## IV. CONCLUSION

This work paved a new way to tackle with ISPs via exploiting model-based network learning strategy. By introducing a simple linear model for ISPs, a new model with network-driven regularizer was proposed. For attaining an efficient end-to-end learning, a network architecture and the estimation of hyper-parameters of the network were presented. Experimental results validated its anti-interference ability, i.e., the robustness of the training procedure. Moreover, reconstruction improvement over the traditional linear algorithm to some extent was demonstrated, i.e., it is superior to the classical linear methods and is comparable to the state-of-the-art nonlinear methods.

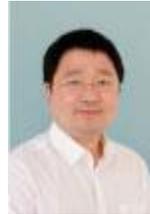

Huilin Zhou was born in Jiangxi, China, in 1979.He received the Ph.D. degree in space physics from Wuhan University, Wuhan, China, in 2006.In 2011, he was a visiting scholar at the State Key Laboratory of Radar Signal Processing at Xidian University. He is currently a Professor with the Cognition Sensor Network Laboratory, School of Information Engineering, Nanchang University, Nanchang, China. His research interests include radar system, radar signal processing, and radar imaging.

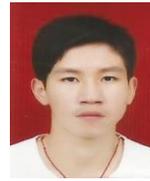

Tao Ouyang was born in 1996 in Jiangxi, China. He is currently pursuing the master's degree in electromagnetic field at Nanchang University. He is mainly engaged in the research of inverse scattering imaging methods and radar signal processing

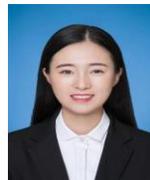

.Yadan Li was born in 1991. She received her bachelor's degree from the school of information engineering, Nanchang University of China.From 2017 to now, she is studying for a master's degree in the school of information engineering, Nanchang University, studying the direction of electromagnetic inverse scattering

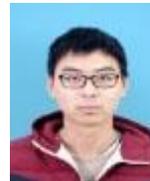

Jian Liu was born in 1995 in Jiangxi, China. He received his bachelor's degree in 2018 at the Department of Electronic Information, School of Information Engineering, Nanchang University. He is currently pursuing the master's degree in electromagnetic field at Nanchang University. He is mainly engaged in the research of inverse scattering imaging methods and radar signal processing.

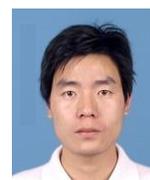

Qiegen Liu (M'16-SM'2019) received the B.S. degree in Applied Mathematics from Gannan Normal University, M. S. degree in Computation Mathematics and Ph.D. degree in Biomedical Engineering from Shanghai Jiaotong University (SJTU). Since 2012, he has been with School of Information Engineering, Nanchang University, Nanchang, P. R. China, where he is currently an Associate Professor. During 2015-2017, he is also a postdoc in UIUC and University of Calgary. His current research interest is sparse representations, deep learning and their applications in image processing, computer vision and MRI reconstruction.